\documentclass[conference]{IEEEtran}

\IEEEoverridecommandlockouts
\ifCLASSINFOpdf
\else
\usepackage[dvips]{graphicx}
\fi

\usepackage{amsmath,amssymb,amsfonts}
\usepackage{array}
\usepackage[caption=false,font=normalsize,labelfont=sf,textfont=sf]{subfig}
\usepackage{textcomp}
\usepackage{url}
\usepackage{verbatim}
\usepackage{graphicx}
\usepackage{cite}
\usepackage{xcolor}
\usepackage{orcidlink}
\usepackage[acronym,shortcuts]{glossaries}
\usepackage{bm}
\usepackage{relsize}
\usepackage{soul}
\usepackage{algorithm}
\usepackage{algcompatible}
\usepackage{mathtools}
\usepackage{bm}
\usepackage{lipsum}
\usepackage{mathtools}
\usepackage{dblfloatfix}
\hypersetup{draft}

\makeatletter
\newcommand\fs@spaceruled{\def\@fs@cfont{\bfseries}\let\@fs@capt\floatc@ruled
  \def\@fs@pre{\vspace{0.4\baselineskip}\hrule height.8pt depth0pt \kern2pt}%
  \def\@fs@post{\kern2pt\hrule\relax}%
  \def\@fs@mid{\kern2pt\hrule\kern2pt}%
  \let\@fs@iftopcapt\iftrue}
\makeatother

\def\BibTeX{{\rm B\kern-.05em{\sc i\kern-.025em b}\kern-.08em
T\kern-.1667em\lower.7ex\hbox{E}\kern-.125emX}}

\newcommand{\trans}[0]{^{\mathsf{T}}}

\newcommand{\herm}[0]{^{\mathsf{H}}}

\newcommand{\Real}[1]{\Re\{{#1}\}}
\newcommand{\Imag}[1]{\Im\{{#1}\}}

\newacronym{CSI}{CSI}{channel state information}
\newacronym{LoS}{LoS}{line-of-sight}
\newacronym{NLoS}{NLoS}{non-LoS}
\newacronym{RPE}{RPE}{radar parameter estimation}
\newacronym{OTFS}{OTFS}{orthogonal time frequency space}
\newacronym{AFDM}{AFDM}{affine frequency division multiplexing}
\newacronym{CRLB}{CRLB}{Cram{\`e}r-Rao lower bound}
\newacronym{BCRLB}{BCRLB}{Bayesian Cram{\`e}r-Rao lower bound}
\newacronym{BBI}{BBI}{Bayesian bilinear inference}
\newacronym{AoA}{AoA}{angle-of-arrival}
\newacronym{SNR}{SNR}{signal-to-noise ratio}
\newacronym{ML}{ML}{maximum likelihood}
\newacronym{MIMO}{MIMO}{multiple-input multiple-output}
\newacronym{SIMO}{SIMO}{single-input multiple-output}
\newacronym{SISO}{SISO}{single-input single-output}
\newacronym{MUSIC}{MUSIC}{multiple signal classification}
\newacronym{MU}{MU}{multi-user}
\newacronym{ROOT-MUSIC}{ROOT-MUSIC}{ROOT multiple signal classification}
\newacronym{JCAS}{JCAS}{joint communication and sensing}
\newacronym{JCR}{JCR}{joint communications and radar}
\newacronym{ISAC}{ISAC}{integrated sensing and communications}
\newacronym{3D}{3D}{three-dimensional}
\newacronym{2D}{2D}{two-dimensional}
\newacronym{1D}{1D}{one-dimensional}
\newacronym{RX}{RX}{receive}
\newacronym{TX}{TX}{transmit}
\newacronym{BF}{BF}{beamformer}
\newacronym{ROI}{ROI}{region of interest}
\newacronym{mmWave}{mmWave}{millimeter-wave}
\newacronym{MF}{MF}{matched-filter}
\newacronym{DD}{DD}{delay-Doppler}
\newacronym{SotA}{SotA}{state-of-the-art}
\newacronym{ULA}{ULA}{uniform linear array}
\newacronym{QAM}{QAM}{quadrature amplitude modulation}
\newacronym{ISFFT}{ISFFT}{inverse symplectic finite Fourier transform}
\newacronym{SFFT}{SFFT}{symplectic finite Fourier transform}
\newacronym{ISI}{ISI}{inter-symbol interference}
\newacronym{AWGN}{AWGN}{additive white Gaussian noise}
\newacronym{MSE}{MSE}{mean-squared-error}
\newacronym{LMMSE}{LMMSE}{linear minimum mean square error}
\newacronym{RMSE}{RMSE}{root mean square error}
\newacronym{ESPRIT}{ESPRIT}{estimation of signal parameters via rotational invariant techniques}
\newacronym{OFDM}{OFDM}{orthogonal frequency division multiplexing}
\newacronym{OCDM}{OCDM}{orthogonal chirp division multiplexing}
\newacronym{BS}{BS}{base station}
\newacronym{UE}{UE}{user equipment}
\newacronym{JCEDD}{JCEDD}{joint channel estimation and data detection}
\newacronym{PDA}{PDA}{probabilistic data association}
\newacronym{PMF}{PMF}{probability mass function}
\newacronym{PBiGaBP}{PBiGaBP}{parametric bilinear Gaussian belief propagation}
\newacronym{PBiGAMP}{PBiGAMP}{parametric bilinear generalized approximate message passing}
\newacronym{GaBP}{GaBP}{Gaussian belief propagation}
\newacronym{FT}{FT}{frequency-time}
\newacronym{DFT}{DFT}{discrete Fourier transform}
\newacronym{IDFT}{IDFT}{inverse discrete Fourier transform}
\newacronym{TD}{TD}{time domain}
\newacronym{wlg}{w.l.g.}{without loss of generality}
\newacronym{CP}{CP}{cyclic prefix}
\newacronym{DAF}{DAF}{discrete affine Fourier}
\newacronym{DAFT}{DAFT}{discrete affine Fourier transform}
\newacronym{IDAFT}{IDAFT}{inverse discrete affine Fourier transform}
\newacronym{CPP}{CPP}{\textit{chirp-periodic} prefix}
\newacronym{IDZT}{IDZT}{inverse discrete Zak transform}
\newacronym{DZT}{DZT}{discrete Zak transform}
\newacronym{P/S}{P/S}{parallel-to-serial}
\newacronym{S/P}{S/P}{serial-to-parallel}
\newacronym{SBL}{SBL}{sparse Bayesian learning}
\newacronym{MPA}{MPA}{message passing algorithms}
\newacronym{EM}{EM}{expectation maximization}
\newacronym{sIC}{soft IC}{soft interference cancellation}
\newacronym{soft RG}{soft RG}{soft replica generation}
\newacronym{BG}{BG}{belief generation}
\newacronym{SGA}{SGA}{scalar Gaussian approximation}
\newacronym{CLT}{CLT}{central limit theorem}
\newacronym{PDF}{PDF}{probability density function}
\newacronym{QPSK}{QPSK}{quadrature phase-shift keying}
\newacronym{ICI}{ICI}{inter-carrier interference}
\newacronym{BER}{BER}{bit error rate}
\newacronym{DoF}{DoF}{degrees-of-freedom}
\newacronym{VGA}{VGA}{vector Gaussian approximation}
\newacronym{FD}{FD}{full-duplex}
\newacronym{SIC}{SIC}{self-interference cancellation}
\newacronym{NMSE}{NMSE}{normalized mean square error}
\newacronym{KL}{KL}{Kullback-Leibler}
\newacronym{LASSO}{LASSO}{least absolute shrinkage and selection operator}
\newacronym{FP}{FP}{fractional programming}
\newacronym{CC}{CC}{communication-centric}
\newacronym{RC}{RC}{raised-cosine}
\newacronym{RRC}{RRC}{root raised-cosine}
\newacronym{6G}{6G}{sixth-generation}
\newacronym{V2X}{V2X}{vehicle-to-everything}
\newacronym{LEO}{LEO}{low-earth orbit}
\newacronym{I/O}{I/O}{input-output}
\newacronym{CE}{CE}{channel estimation}

\begin{document}
\title{Low Complexity Joint Channel Estimation and Data Detection for AFDM Receivers With Oversampling}
\author{\IEEEauthorblockN{Kuranage Roche Rayan Ranasinghe*\textsuperscript{\orcidlink{0000-0002-6834-8877}}, Yao Ge$^\dag$\textsuperscript{\orcidlink{0000-0002-3293-2051}}, Giuseppe Thadeu Freitas de Abreu*\textsuperscript{\orcidlink{0000-0002-5018-8174}} and Yong Liang Guan$^\ddag$\textsuperscript{\orcidlink{0000-0002-9757-630X}}}
\IEEEauthorblockA{\textit{*School of Computer Science and Engineering, Constructor University, 28759 Bremen, Germany} \\ \textit{$^\dag$Continental-NTU Corporate Lab, Nanyang Technological University, 639798, Singapore} \\ \textit{$^\ddag$School of Electrical and Electronic Engineering, Nanyang Technological University, 639798, Singapore} \\
(kranasinghe,gabreu)@constructor.university, (yao.ge,eylguan)@ntu.edu.sg\\[-3ex]}
}
\maketitle 

\begin{abstract}
In this paper, we propose a novel low complexity \ac{TD} oversampling receiver framework under \ac{AFDM} waveforms for \ac{JCEDD}.
Leveraging a generalized doubly-dispersive channel model, we first derive the \ac{I/O} relationship for arbitrary waveforms when oversampled in the \ac{TD} and present the \ac{I/O} relationship for \ac{AFDM} as an example.
Subsequently, utilizing the multiple sample streams created via the oversampling procedure, we use the \ac{PBiGaBP} technique to conduct \ac{JCEDD} for decoding the transmitted data and estimating the complex channel coefficients.
Simulation results verify significant performance improvements both in terms of data decoding and  complex channel coefficient estimation with improved robustness against a varying number of pilots over a conventional Nyquist sampling rate receiver.
\end{abstract}


\glsresetall

\vspace{-1ex}
\section{Introduction}
\vspace{-1ex}

% Need for high performance in high-mobility scenarios
High-mobility scenarios have been a capitalizing factor driving the paradigm-shifting research for enabling the envisioned \ac{6G} wireless network \cite{WuACCESS2016,JiangOJCom2021}, with many extended applications such as \ac{V2X} communications \cite{ChenCSM2017} and \ac{LEO} satellite networks \cite{ShiNetwork2024}.
However, the consequent fast-varying surrounding gives rise to large Doppler shifts which are commonly modelled using the doubly-dispersive channel structure\cite{Rou_SPM_2024}.

% Intro. to waveforms in use (include examples from OFDM and OTFS)
Conventionally, the use of \ac{OFDM} waveforms in doubly-dispersive environments results in severe degradation of communications performance \cite{WangTWC2006} due to the presence of \ac{ICI}.
To combat this, the \ac{OTFS} waveform \cite{Hadani_WCNC_2017} multiplexes information symbols in the \ac{DD} domain giving it inherent robustness against \ac{ICI} due to the additional \ac{DoF} exploited.
However, due to its \ac{2D} structure, \ac{OTFS} incurs a much higher modulation complexity compared to \ac{OFDM}.
As a result, the \ac{AFDM} waveform \cite{Bemani_TWC_2023} retains the same communication performance with a lower pilot overhead compared to \ac{OTFS} while achieving full diversity without the higher modulation complexity due to its \ac{1D} structure, making it a potential candidate for high-mobility communications.

% Intro. to oversampling (include a bunch of citations) and JCDE at the Reciever (point out how current studies assume full channel knowledge)
Therefore, motivated by previous studies on oversampling receivers for the \ac{OFDM} \cite{ShiTVT2010,WuTCom2011} and \ac{OTFS} \cite{GeTWC2021,PriyaTWC2024} waveforms which can further improve communication performance and robustness against fractional Doppler shifts, we consider an oversampling receiver for \ac{AFDM} waveforms in high-mobility surroundings.
In contrast to all the aforementioned literature which only considered data decoding with full knowledge of the complex channel coefficients (which is usually not the case) at the receiver, we now consider the \ac{JCEDD} problem, and solve it via the \ac{PBiGaBP} algorithm to both decode the information symbols and estimate the complex channel coefficients.

% Describe our contributions
Summarizing our main contributions, we first formulate the \ac{I/O} relationship under a doubly-dispersive channel structure for arbitrary oversampled \ac{TD} signals which is then presented for \ac{AFDM}.
Next, using the multiple sample streams generated by the oversampling procedure, we conduct \ac{JCEDD} via the \ac{PBiGaBP} mechanism to extract both payload data and the complex channel coefficients with a low complexity.
Finally, simulation results are presented which demonstrate a significant performance gain when an oversampled \ac{PBiGaBP} receiver -- which takes advantage of the larger system size due to oversampling in the belief generation stage -- is used instead of a typical Nyquist rate \ac{PBiGaBP} receiver.
In addition, it can also be observed that the oversampling procedure increases robustness against a varying number of pilots due to the increase in factor nodes of the overall \ac{PBiGaBP} framework.

\section{Generalized Oversampled \ac{TD} \ac{I/O} Relationship}
\label{sec:system_model}
\vspace{-0.5ex}

\subsection{Doubly-Dispersive Channel Model}
\label{subsec:DD_channel_model}
\vspace{-1ex}

\begin{figure*}[!b]
    \vspace{-3ex}
    \hrulefill
    \setcounter{equation}{6}
    \normalsize
    \begin{equation}
    \label{eq:diagonal_CP_matrix_def}
    {
    \mathbf{\Phi}_p \triangleq \text{diag}\Big( [ \overbrace{e^{-j2\pi\phi_\mathrm{CP}(\ell_p)}, e^{-j2\pi\phi_\mathrm{CP}(\ell_p-1)}, \dots, e^{-j2\pi\phi_\mathrm{CP}(2)}, e^{-j2\pi\phi_\mathrm{CP}(1)}}^{\ell_p \; \text{terms}}, \overbrace{1, 1, \dots, 1, 1}^{N - \ell_p \; \text{ones}}] \Big) \in \mathbb{C}^{N \times N}.
    \vspace{-5ex}}
    \end{equation}
    
    % NOTE: Phi_p is an identity matrix for OFDM and OTFS
    %
    \setcounter{equation}{7}
    \begin{equation}
    \label{eq:diagonal_Doppler_matrix_def}
    \boldsymbol{\Omega}^{(g)} \triangleq \text{diag}\Big([e^{-j2\pi g /(NG)},e^{-j2\pi (g+G) /(NG)},\dots,e^{-j2\pi \big(g+(N-2)G\big) /(NG)}, e^{-j2\pi \big(g+(N-1)G\big) /(NG)}]\Big) \in \mathbb{C}^{N \times N}.
    \end{equation}
    \setcounter{equation}{0}
    \vspace{-3ex}
\end{figure*}

Utilizing the doubly-dispersive wireless channel {\cite{Bliss_Govindasamy_2013,Rou_SPM_2024}} to model the surrounding with one \ac{LoS} and $P$ \ac{NLoS} propagation paths corresponding to the $P$ scatterers in the surrounding yields the \ac{TD} channel impulse response function $h(t, \tau)$ in the continuous time-delay domain given by
\begin{equation}
\label{eq:doubly_dispersive_time_delay_channel}
h(t,\tau) \triangleq \sum_{p=0}^P h_p \cdot e^{j2\pi \nu_p t} \cdot \delta(\tau - \tau_p),
\end{equation}
where $p = 0$ denotes the \ac{LoS} path and $p \in \{1,\cdots\!,P\}$ corresponds to the \ac{NLoS} path from each $p$-th scatterer; 
$h_p \in \mathbb{C}$ is the $p$-th complex fading channel coefficient;
$\tau_p \in [0,\tau_\text{max}]$ is the $p$-th path delay bounded by the maximum delay $\tau_\text{max}$ and 
$\nu_p \in [-\nu_\text{max},\nu_\text{max}]$ is the $p$-th Doppler shift bounded by the maximum Doppler shift $\nu_\text{max}$.

\subsection{Input-Output Relationship with TD Oversampling}
\label{subsec:IO_TD_oversampling}

Given an arbitrary baseband \ac{TD} transmit signal $s(t)$ with bandwidth $B$, the received signal $r(t)$ after transmission over the doubly-dispersive channel $h(t,\tau)$ given in equation \eqref{eq:doubly_dispersive_time_delay_channel} can be described in terms of their linear convolution \cite{Bliss_Govindasamy_2013,Rou_SPM_2024}
\begin{eqnarray}
    \label{eq:time-domain-io_relationship}
    r(t) = s(t) * h(t,\tau) + w(t)  && \\
    &&\hspace{-29ex} \triangleq \int_{-\infty}^{+\infty} \! s(t-\tau) \bigg( \sum_{p=0}^P h_p \cdot e^{j2\pi \nu_p t} \cdot \delta(\tau - \tau_p) \bigg) d\tau + w(t), \nonumber
\end{eqnarray}
where $w(t)$ is the \ac{TD} \ac{AWGN}.

Next, oversampling the \ac{TD} received signal in equation \eqref{eq:time-domain-io_relationship} at an arbitrary sampling frequency $F_\mathrm{S} \triangleq G \cdot f_\mathrm{s} = \frac{G}{T_\mathrm{S}}$ where $G$ is an integer oversampling factor ($i.e.,$ each symbol is sampled $G$ times), $f_\mathrm{S} \triangleq B$ is the typical Nyquist sampling rate (also known as the symbol-spaced sampling rate \cite{ShiTVT2010,GeTWC2021,PriyaTWC2024}) and $T_\mathrm{S}$ is the corresponding Nyquist sampling period yields \cite{Rou_SPM_2024}
%
% {\color{blue}
% \begin{align}
%     \label{eq:oversampled_TD_IO_relationship}
%     r(n T_\mathrm{S} + \frac{g}{G} T_\mathrm{S}) &= \Bigg[ \sum_{\ell = 0}^\infty s(n T_\mathrm{S} - \ell T_\mathrm{S}) \\ 
%     &\hspace{-12ex}\times \bigg( \sum_{p=0}^P h_p \cdot e^{j2\pi \nu_p (n + \frac{g}{G}) T_\mathrm{S}} \cdot \text{sinc}\big(\ell - \frac{nG}{nG + g} \cdot \frac{\tau_p}{T_\mathrm{S}}\big) \bigg) \Bigg]\nonumber \\ 
%     &\hspace{-12ex}+ w(n T_\mathrm{S} + \frac{g}{G} T_\mathrm{S}), \hspace{16ex} g = 0,\cdots,G-1, \nonumber
% \end{align}
% }
%
% \begin{align}
%     \label{eq:oversampled_TD_IO_relationship}
%     r(n T_\mathrm{S} + \frac{g}{G} T_\mathrm{S}) &= \Bigg[ \sum_{\ell = 0}^\infty s(n T_\mathrm{S} - \ell T_\mathrm{S}) \\ 
%     &\hspace{-12ex}\times \bigg( \sum_{p=0}^P h_p \cdot e^{j2\pi \nu_p n \frac{T_\mathrm{S}}{(g+1)}} \cdot \text{sinc}\Big(\ell - \frac{(g+1) \tau_p}{T_\mathrm{S}}\Big) \bigg) \Bigg]\nonumber \\ 
%     &\hspace{-12ex}+ w(n T_\mathrm{S} + \frac{g}{G} T_\mathrm{S}), \hspace{12ex} g = 0,\cdots,G-1, \nonumber
% \end{align}
\begin{align}
    \label{eq:oversampled_TD_IO_relationship}
    r(n T_\mathrm{S} + \tfrac{g}{G} T_\mathrm{S}) &= \Bigg[ \sum_{\ell = 0}^\infty s(n T_\mathrm{S} + \tfrac{g}{G} T_\mathrm{S}  - \ell T_\mathrm{S} - \tfrac{g}{G} T_\mathrm{S}) \nonumber\\ 
    &\hspace{-13.5ex}\times \bigg( \sum_{p=0}^P h_p \cdot e^{j2\pi \nu_p (n + \frac{g}{G}) T_\mathrm{S}} \cdot \text{P}_\mathrm{RC}\big(\ell T_\mathrm{S} \!+\! \tfrac{g}{G} T_\mathrm{S} \!-\! \tau_p \!-\! \tfrac{g}{G} T_\mathrm{S}\big) \bigg) \Bigg] \nonumber \\
    &\hspace{-13.5ex}+ w(n T_\mathrm{S} + \tfrac{g}{G} T_\mathrm{S}), 
\end{align}
where $n \in \{ 0,\cdots,N-1 \}$ and $\ell \in \{ 0,\cdots,N-1 \}$ denote the discrete time and delay indices, respectively; $\text{P}_\mathrm{RC}$ is the \ac{RC} rolloff pulse if the transmit filter response is a \ac{RRC} rolloff pulse and the receive filter is its corresponding matched filter; $g \in \{ 
0,\cdots,G-1 \}$ refers to the $g$-th sample stream due to oversampling and $w(n T_\mathrm{S} + \tfrac{g}{G} T_\mathrm{S}), \forall g$ is correlated due to the \ac{RC} filter adopted (refer to Section \ref{subsec:noise_correlation} for more details on the correlation).

Since the Nyquist frequency rate $f_\mathrm{S}$ in typical wideband communication systems is already sufficiently high to approximate the normalized path delays $\ell_p \!\triangleq\! \tau_p f_\mathrm{S} \!=\! \frac{\tau_p}{T_\mathrm{S}} \in [ 0,\ell_\text{max} ]$ to the nearest integer with negligible error ($i.e.,$ $\ell_p \!-\! \lfloor \frac{\tau_p}{T_\mathrm{S}} \rfloor \!\approx\! 0$) and the normalized digital Doppler shift of the $p$-th path is given by $f_p \!\triangleq\! \frac{N \nu_p}{f_\mathrm{S}} \!=\! N \nu_p T_\mathrm{S} \!\in\! [-f_\text{max},\!f_\text{max}]$ \cite{Rou_SPM_2024}, the \ac{RC} pulses become equivalent to unit impulses due to the integer delay such that the $g$-th corresponding sampled sequences become
\vspace{-1ex}
\begin{equation}
    \label{eq:oversampled_sequences}
    r^{(g)}[n] \!=\!\! \sum_{\ell=0}^\infty \!s[n - \ell] \!\bigg( \sum_{p=0}^P \!h_p \cdot e^{j2\pi f_p (\!\frac{g + nG}{NG}\!)} \!\cdot \delta[\ell - \ell_p] \!\bigg) \!\!+\! w^{(g)}[n],
\end{equation}
where $r^{(g)}[n]$, $s[n]$\footnote{Note that in contrast to $r^{(g)}[n]$ and $w^{(g)}[n]$, $s[n]$ does not depend on the $g$-th oversampled instance since the same symbol is sampled multiple times during the oversampling procedure.} and $w^{(g)}[n]$ are the $g$-th oversampled sequences of $r(t)$, $s(t)$ and $w(t)$, respectively with $\delta[\,\cdot\,]$ defined to be the discrete unit impulse function.

Incorporating a \ac{CP} of length $N_\mathrm{CP}$ samples such that $N_\mathrm{CP} \geq \ell_\text{max}$ yields
\begin{equation}
    \label{eq:cyclic_prefix_def}
    s[n'] = s[N + n'] \cdot e^{j2\pi \phi_\mathrm{CP}(n')},
\end{equation}
where $n' \in \{ -1,\cdots,-N_\mathrm{CP} \}$ and $\phi_\mathrm{CP}(n')$ is a waveform specific multiplicative phase term.

Finally, using the \ac{CP} definition in equation \eqref{eq:cyclic_prefix_def} to process the linear convolution after the removal of the \ac{CP} parts yields the circular convolutional form \cite{Rou_SPM_2024}
\vspace{-1ex}
\begin{equation}
\label{eq:channel_matrix_TD_general}
\mathbf{r}^{(g)} \!=\! \bigg( \sum_{p=0}^P h_p \cdot \mathbf{\Phi}_p 
\cdot\! \big(\!\boldsymbol{\Omega}^{(g)}\!\big)^{\!f_p} \!\cdot \mathbf{\Pi}^{\ell_p} \!\bigg) \cdot \mathbf{s} + \mathbf{w}^{(g)} \!=\! \bm{\Psi}^{(g)} \cdot \mathbf{s} + \mathbf{w}^{(g)}\!,\!\!\!
\vspace{-1ex}
\end{equation}
where $\mathbf{s} \in \mathbb{C}^{N \times 1}$, $\mathbf{r}^{(g)} \in \mathbb{C}^{N \times 1}$ and $\mathbf{w}^{(g)} \in \mathbb{C}^{N \times 1}$ are the transmit, received and \ac{AWGN} signal vectors consisting of $N$ samples, respectively;
%
%It was shown in  that for two arbitrary time domain signals $s(t)$ and $r(t)$, with their sampled\footnote{At a sampling rate $f_\mathrm{S} \triangleq \frac{1}{T_\mathrm{S}}$[Hz], where $T_\mathrm{S}$ denotes the delay resolution and sampling interval such that $\tau = \ell\cdot T_\mathrm{S}$, with $\ell$ defined to be the discrete delay index.} versions given by $r[n]$ and $s[n]$, with $n \in \{ 0,\dots,N-1 \}$, the vectorized input-output relationship after convolution with the doubly dispersive channel described in eq. \eqref{eq:doubly_dispersive_time_delay_channel} can be represented by
%
$\bm{\Psi}^{(g)} \in \mathbb{C}^{N \times N}$ is the effective circular convolutional channel matrix, $\mathbf{\Phi}_p \in \mathbb{C}^{N \times N}$ described in equation \eqref{eq:diagonal_CP_matrix_def} is a diagonal matrix capturing the effect of the \ac{CP} phase with $\phi_\mathrm{CP}(n)$ denoting the waveform-dependent phase function \cite{Rou_SPM_2024} on the sample index $n \in \{0,\cdots,N-1\}$;
$\boldsymbol{\Omega}^{(g)} \!\in\! \mathbb{C}^{N \times N}$ described in equation \eqref{eq:diagonal_Doppler_matrix_def} is a diagonal matrix dependent on both the $g$-th sampled instance and the oversampling factor $G$;
and $\mathbf{\Pi}\in \{0,1\}^{N \times N}$ is the forward cyclic shift matrix, with elements given by
\vspace{-0.5ex}
\setcounter{equation}{8}
\begin{equation}
\label{eq:PiMatrix}
\pi_{i,j} = \delta_{i,j+1} + \delta_{i,j-(N-1)}\;\; \text{where}\;\; \delta _{i,j} \triangleq
\begin{cases}
0 & \text{if }i\neq j,\\
1 & \text{if }i=j.
\end{cases}
\end{equation}

%------------------------------------------------------------------
% \begin{figure*}[b]
% % \hrulefill
% \setcounter{equation}{9}
% \normalsize
% \begin{equation}
% \label{eq:AFDM_diagonal_CP_matrix_def}
% \bm{\varPhi}_p \triangleq \text{diag}\Big( [ \overbrace{e^{-j2\pi c_1 (N^2-2N\ell_p)}, e^{-j2\pi c_1 (N^2-2N(\ell_p-1))}, \cdots, e^{-j2\pi c_1 (N^2-2N)}}^{\ell_p \; \text{terms}}, \overbrace{1, 1, \dots, 1, 1}^{N - \ell_p \; \text{ones}}] \Big) \in \mathbb{C}^{N \times N}.
% \vspace{-1ex}
% \end{equation}
% \setcounter{equation}{5}
% \end{figure*}
%------------------------------------------------------------------

Furthermore, the roots-of-unity matrix $\boldsymbol{\Omega}^{(g)}$ and the forward cyclic shift matrix $\mathbf{\Pi}$ are respectively exponentiated\footnote{Matrix exponentiation of $\boldsymbol{\Omega}^{(g)}$ is equivalent to an element-wise exponentiation of the diagonal entries, and the matrix exponentiation of $\mathbf{\Pi}^k$ is equivalent to a forward (left) circular shift operation of $k$ indices.} to the power of $f_p$ and $\ell_p$, which are the normalized digital Doppler frequency and the normalized delay of the $p$-th path.

\subsection{Oversampled Noise Correlation Discussion}
\label{subsec:noise_correlation}

As previously discussed briefly in Section \ref{subsec:IO_TD_oversampling}, the \ac{TD} \ac{AWGN} vectors $\mathbf{w}^{(g)}, \forall g$ given in equation \eqref{eq:channel_matrix_TD_general} are correlated as a result of oversampling each symbol \cite{StevenBOOK,GeTWC2021,PriyaTWC2024}.

% {\color{red}How can we detail the procedure you use in your paper for correlating the noise samples?}

Following \cite{PriyaTWC2024}, we denote $\mathbf{C}_\mathbf{w} \in \mathbb{C}^{G \times G}$ to be the covariance matrix of the oversampled discrete \ac{TD} \ac{AWGN} vectors $\mathbf{w}^{(g)}, \forall g$ defined as
\begin{equation}
    \label{eq:noise_corr_matrix}
    \mathbf{C}_\mathbf{w} \triangleq \text{Toepl}\Big[ c_w(\frac{0}{G}), \dots, c_w(\frac{g}{G}), \dots, c_w(\frac{G-1}{G}) \Big],
\end{equation}
with $c_w(\frac{g}{G}) \triangleq \frac{\sigma_w^2}{2} \text{P}_\mathrm{RC}(\frac{g}{G})$ where $\sigma_w^2$ is the \ac{AWGN} variance and $\text{Toepl}(\mathbf{a})$ denotes the Toeplitz matrix with elements $\mathbf{a}$. 

Next, a Cholesky decomposition of $\mathbf{C}_\mathbf{w} = \mathbf{L}_\mathbf{C}\herm \cdot \mathbf{L}_\mathbf{C}$ yields the correlation shaping matrix $\mathbf{L}_\mathbf{C} \in \mathbb{C}^{G \times G}$.

Finally, denoting and stacking a set of $G$ uncorrelated \ac{AWGN} noise vectors $\bm{w}_\mathrm{UC} \!\triangleq\! [ \mathbf{w}_\mathrm{UC}^{(0)}, \dots, \!\mathbf{w}_\mathrm{UC}^{(g)}, \dots, \!\mathbf{w}_\mathrm{UC}^{(G-1)} ]\trans \in \mathbb{C}^{G \times N}$ yields the stacked correlated \ac{AWGN} noise vectors given in equation \eqref{eq:channel_matrix_TD_general} to be
\begin{equation}
    \label{eq:correlated_noise_vectors}
    \mathbf{L}_\mathbf{C} \cdot \bm{w}_\mathrm{UC} \triangleq [ \mathbf{w}^{(0)}, \dots, \mathbf{w}^{(g)}, \dots, \mathbf{w}^{(G-1)} ]\trans \in \mathbb{C}^{G \times N}.
\end{equation}

\section{Oversampled \ac{AFDM} \ac{I/O} Relationship}

We consider a typical point-to-point single antenna communication system using \ac{AFDM} waveforms \cite{Bemani_TWC_2023,RanasingheARXIV2024} composed of a transmitter and an oversampling-enabled receiver, with $P$ significant scatterers in the environment.
Using the general derivation presented in Section \ref{sec:system_model} for an arbitrary \ac{TD} transmit signal $s(t)$, henceforth \ac{wlg}, we consider the \ac{AFDM} waveform, keeping in mind that the same logic also applies to \ac{OFDM} and \ac{OTFS} waveforms.

Let $\mathbf{x} \in \mathbb{C}^{N \times 1}$ denote the information vector with elements drawn from an arbitrary complex digital constellation $\mathcal{C}$, with cardinality $Q \triangleq |\mathcal{C}|$ and average symbol energy $\sigma^2_X$.
The corresponding \ac{AFDM} modulated transmit signal of $\mathbf{x}$ is given by its \ac{IDAFT}, $i.e.,$
\begin{equation}
\label{eq:AFDM_moduation}
\mathbf{s}_\text{AFDM} = (\mathbf{\Lambda}_1\herm \mathbf{F}_{N}\herm \mathbf{\Lambda}_2\herm) \cdot \mathbf{x} \in \mathbb{C}^{N \times 1},
\end{equation}
where $\mathbf{F}_N \in \mathbb{C}^{N \times N}$ denotes the $N$-point normalized \ac{DFT} matrix and the two diagonal chirp matrices $\mathbf{\Lambda}_i \!\in\! \mathbb{C}^{N \times N}$ are defined as
\begin{equation}
\label{eq:lambda_def}
\!\!\!\!\mathbf{\Lambda}_i \!\triangleq\! \mathrm{diag}\Big(\big[1, \cdots\!, e^{-j2\pi c_i n^2}\!\!, \cdots\!, e^{-j2\pi c_i (N-1)^2}\big]\Big),
\end{equation}
where the first central frequency $c_1$ is selected for optimal robustness to doubly-dispersivity based on the channel statistics \cite{Bemani_TWC_2023}\footnote{Note that there are no changes to $f_p$ due to the oversampling factor $G$ causing no change to the corresponding effective value of $c_1$ and the \ac{AFDM} modulation procedure.}, and the second central frequency $c_2$ which does not depend on $g$ can be exploited for waveform design and applications \cite{zhu2023low,rou2024afdm}.

In addition, the \ac{AFDM} modulated signal also requires the insertion of a \ac{CPP} to mitigate the effects of multipath propagation \cite{Bemani_TWC_2023} analogous to the \ac{CP} in \ac{OFDM}, whose multiplicative phase function for equation \eqref{eq:diagonal_CP_matrix_def} is given by $\phi_\mathrm{CPP}(n) = c_1(g) (N^2 - 2Nn)$ \cite{Rou_SPM_2024}.
Correspondingly, the received \ac{AFDM} signal vector is given by
\begin{equation}
\label{eq:AFDM_received vector_in_TD}
\mathbf{r}_\text{AFDM}^{(g)} \triangleq \bm{\Psi}^{(g)} \cdot \mathbf{s}_\text{AFDM} + \mathbf{w}^{(g)} \in \mathbb{C}^{N\times 1}.
\end{equation}

Then, the received signal in equation \eqref{eq:AFDM_received vector_in_TD} is demodulated via the \ac{DAFT} to yield
\begin{align}
\mathbf{y}_\text{AFDM}^{(g)} \!&=\! (\mathbf{\Lambda}_2 \mathbf{F}_{N} \mathbf{\Lambda}_1) \cdot \mathbf{r}_\text{AFDM}^{(g)} \in \mathbb{C}^{N\times 1} \nonumber \\
&\hspace{-5ex}= \! (\mathbf{\Lambda}_2 \mathbf{F}_{N} \mathbf{\Lambda}_1)\! \cdot \!\bigg(\!\!~\! \sum_{p=0}^P h_p \!\!\cdot\!\bm{\Phi}_p\! \!\cdot\! \big(\boldsymbol{\Omega}^{(g)}\big)^{\!f_p} \!\!\cdot\! \mathbf{\Pi}^{\ell_p}\!\!\bigg)\! \cdot \!(\mathbf{\Lambda}_1\herm \mathbf{F}_{N}\herm \mathbf{\Lambda}_2\herm) \!\cdot\! \mathbf{x} \nonumber \\
&\hspace{-4.9ex}+ (\mathbf{\Lambda}_2 \mathbf{F}_{N} \mathbf{\Lambda}_1)\mathbf{w}^{(g)}. \label{eq:AFDM_demodulation}
\end{align}

In light of the above, the final oversampled \ac{I/O} relationship of \ac{AFDM} over doubly-dispersive channels is given by
\begin{equation}
\mathbf{y}_\text{AFDM}^{(g)} = \mathbf{G}_\text{AFDM}^{(g)} \cdot \mathbf{x} + \tilde{\mathbf{w}}_\text{AFDM}^{(g)} \in \mathbb{C}^{N\times 1},
\label{eq:DAF_input_output_relation}
\end{equation}
where $\tilde{\mathbf{w}}_\text{AFDM}^{(g)} \triangleq (\mathbf{\Lambda}_2 \mathbf{F}_{N} \mathbf{\Lambda}_1)\mathbf{w}^{(g)} \in \mathbb{C}^{N\times 1}$ is an equivalent \ac{AWGN} vector with the same statistical properties\footnote{This is because the \ac{DAFT} is a unitary transformation \cite{Bemani_TWC_2023}.} as $\mathbf{w}^{(g)}$, and $\mathbf{G}_\text{AFDM}^{(g)} \in \mathbb{C}^{N\times N}$ is the effective \ac{AFDM} channel
\begin{align}
\label{eq:DAF_domain_effective_channel}
\!\!\!\!\mathbf{G}_\text{AFDM}^{(g)}\! \triangleq\! \sum_{p=0}^P \!h_p \underbrace{ \!\cdot (\mathbf{\Lambda}_2 \mathbf{F}_{N} \mathbf{\Lambda}_1) \!\!\cdot\!\! \big(\bm{\Phi}_p\! \!\cdot\! \big(\boldsymbol{\Omega}^{(g)}\big)^{\!f_p} \!\!\cdot\!\! \mathbf{\Pi}^{\ell_p}\!\big) \!\!\cdot\!\! (\mathbf{\Lambda}_1\herm \mathbf{F}_{N}\herm \mathbf{\Lambda}_2\herm) }_{\mathbf{\Gamma}_p^{(g)}}.
\end{align}

\section{Joint Channel Estimation and Data Detection}

In this section, the proposed \ac{JCEDD} technique, termed \ac{PBiGaBP}, is introduced for \ac{AFDM} systems, under the general system model described in the previous section.
For later convenience and effective manipulation, we now stack each $g$-th oversampled instance, similar to a \ac{SIMO} scenario \cite{GeTWC2021}, and  express the effective channel given in equation \eqref{eq:DAF_domain_effective_channel} in the form
\begin{equation}
\label{eq:final_generalized_input_output_relation}
\mathbf{y} \!=\!
\begin{bmatrix}
\mathbf{y}^{(0)}\\
\vdots\\
\mathbf{y}^{(g)}\\
\vdots\\
\mathbf{y}^{(G-1)}
\end{bmatrix}
\!=\! \sum_{p=0}^P h_p \cdot
\underbrace{\begin{bmatrix}
\mathbf{\Gamma}_p^{(0)}\\
\vdots\\
\mathbf{\Gamma}_p^{(g)}\\
\vdots\\
\mathbf{\Gamma}_p^{(G-1)}
\end{bmatrix}}_{\mathbf{\Gamma}_p \in \mathbb{C}^{GN \times N}}
\cdot \mathbf{x} + 
\underbrace{\begin{bmatrix}
\tilde{\mathbf{w}}^{(0)}\\
\vdots\\
\tilde{\mathbf{w}}^{(g)}\\
\vdots\\
\tilde{\mathbf{w}}^{(G-1)}
\end{bmatrix}}_{\tilde{\mathbf{w}} \in \mathbb{C}^{GN \times 1}}
\!\in \mathbb{C}^{GN\times 1},
\end{equation}
with the waveform-specific subscript dropped for brevity and where the matrices $\mathbf{\Gamma}_p$ captures the long-term delay-Doppler statistics of the channel, which are assumed to be known\footnote{According to \cite{MatzTWC2005,MishraTWC2022,YangTVT2024}, the delay and Doppler shifts can be assumed constant throughout multiple frame transmissions, enabling prior estimation of $\mathbf{\Gamma}_p, \forall p$ via \ac{RPE} schemes using pilots \cite{RanasingheARXIV2024}.}.

Then, from equation \eqref{eq:final_generalized_input_output_relation}, each received symbol in $\mathbf{y}$ can be described element-wise by
\begin{equation}
y_n = \sum_{p=0}^P \sum_{m=0}^{N-1} h_p \cdot \gamma_{p:n,m} \cdot x_m + \tilde{w}_n, \;\; n = 1,\cdots,GN,
\label{eq:elementwise_final_input_output_relation}
\end{equation}
where we slightly modify the notation formerly employed in equation \eqref{eq:oversampled_sequences} by denoting the $m$-th information symbol as $x_m$ instead of $x[m]$, and accordingly the $n$-th receive signal sample by $y_n$ instead of $y[n]$ for future clarity.

\ac{JCEDD} via \ac{PBiGaBP} was first introduced for \ac{AFDM} systems in \cite{RanasingheARXIV2024}, where it was shown to outperform other alternatives like \ac{OFDM} and \ac{OTFS} in doubly-dispersive channels. 
In this paper, we show that a similar derivation can be used for oversampled \ac{AFDM} systems with pertinent modifications to allow for the increased number of variable nodes.

In this manuscript, we utilize a typical \ac{AFDM} frame structure, with $N_\mathrm{P}$ pilot symbols followed by $N-N_\mathrm{P}$ data symbols, as opposed to the structure with a single pilot followed by a null-guard interval considered in \cite{Bemani_WCL_2024} since we do not consider a monostatic sensing scenario.

The main signal processing operations for the \ac{PBiGaBP} framework used in \ac{AFDM} systems is succinctly provided below.
Note that all the equations are derived for a given $i$-th iteration of the algorithm.

\subsubsection{Soft IC}

For the $n$-th receive signal $y_n$ given in equation \eqref{eq:elementwise_final_input_output_relation}, the soft replicas of the $m$-th symbol and the $p$-th channel gain at the $(i-1)$-th iteration of the algorithm are given by $\hat{x}_{n,m}^{(i-1)}$ and $\hat{h}_{n,p}^{(i-1)}$, respectively.
With the soft replicas in hand, the corresponding \acp{MSE} are given by
\begin{subequations}
\begin{equation}
\hat{\sigma}^{2(i)}_{x:{n,m}} \triangleq \mathbb{E}_{x} \big[ | x - \hat{x}_{n,m}^{(i-1)} |^2 \big]= E_\mathrm{S} - |\hat{x}_{n,m}^{(i-1)}|^2, \forall (n,m),
\label{eq:MSE_x_m}
\end{equation}
\begin{equation}
\hat{\sigma}^{2(i)}_{h:{n,p}} \triangleq \mathbb{E}_{h} \big[ | h - \hat{h}_{n,p}^{(i-1)} |^2 \big] = \sigma_h^2 -|\hat{h}_{n,p}^{(i-1)} |^2, \forall (n,p),
\label{eq:MSE_h_p}
\end{equation}
where $\mathbb{E}_{x}$ denotes the expectation over the all possible symbols $x$ in the constellation $\mathcal{C}$, while $\mathbb{E}_{h}$ denotes the expectation over all possible outcomes of $h\sim\mathcal{CN}(0,\sigma^2_h)$ \cite{RanasingheARXIV2024}, respectively.
\end{subequations}

Next, proceeding with the \ac{sIC} step to compute the symbol- and channel-centric replicas $\tilde{y}_{x:{m,n}}^{(i)}$ and $\tilde{y}_{h:{p,n}}^{(i)}$ and the corresponding variances $\tilde{\sigma}^{2(i)}_{x:{m,n}}$ and $\tilde{\sigma}^{2(i)}_{h:{p,n}}$, equation \eqref{eq:elementwise_final_input_output_relation} straightforwardly gives us
\begin{equation}
\label{eq:soft_IC_process}
\tilde{y}_{x:{m,n}}^{(i)} = y_n - \sum_{p=0}^P \sum_{q \neq m}^{N-1} \hat{h}_{n,p}^{(i-1)} \cdot \gamma_{p:n,q} \cdot \hat{x}_{n,q}^{(i-1)}.
\end{equation}

Exploiting the fact that $\tilde{y}_{x:{m,n}}^{(i)}, \forall m,n$ follow Gaussian \acp{PDF} and defining $\text{y}_n, \forall n$ to be an auxiliary variable, the \ac{sIC} symbol replicas are defined as
\begin{equation}
\tilde{y}_{x:{m,n}}^{(i)} \sim p_{\text{y}_n | x_m}(\text{y}_n | x_m) \propto \text{exp}\bigg(\! - \frac{|\text{y}_n\! -\! \tilde{\gamma}_{x:{m,n}}^{(i)} x_m|^2}{\tilde{\sigma}^{2(i)}_{x:{m,n}}} \bigg),
\label{eq:conditional_PDF_y_nm_x}
\end{equation}
with $\tilde{\gamma}_{x:{m,n}}^{(i)}, \forall m,n$ implicitly denoting the corresponding \ac{sIC} effective channel gains given by
\begin{equation}
\label{eq:effective_channel_gain_a_in_PDF}
\tilde{\gamma}_{x:{m,n}}^{(i)} \triangleq \sum_{p=0}^P \hat{h}_{n,p}^{(i-1)} \gamma_{p:n,m},
\end{equation}
while the \ac{sIC} conditional variances $\tilde{\sigma}^{2(i)}_{x:{m,n}}$ are approximated by replacing the instantaneous values with the long-term statistics as
\vspace{-1ex}
\begin{eqnarray}
\vspace{-1ex}
\hspace{-0.5ex}\tilde{\sigma}^{2(i)}_{x:{m,n}} \triangleq \mathbb{E}_{x,h} \big[ | \tilde{y}_{x:{m,n}}^{(i)} - \tilde{\gamma}_{x:{m,n}}^{(i)} x_m |^2 \big]  && \nonumber \\
&&\hspace{-39.4ex} \approx \sum_{p=0}^P \hat{\sigma}^{2(i-1)}_{h:{n,p}} |\hat{y}^{(i)}_{x_m:n,p}|^2 + \sum_{q \neq m}^{N-1} \hat{\sigma}_{x:{n,q}}^{2(i-1)} |\tilde{\gamma}_{x:q,n}^{(i)}|^2 + N_0 + \nonumber \\
&&\hspace{-40.5ex} \sum_{p=0}^P \!\hat{\sigma}^{2(i-1)}_{h:{n,p}} \!\sum_{q \neq m}^{N-1} \!\hat{\sigma}_{x:{n,q}}^{2(i-1)} |\gamma_{p:n,q}|^2\! +\! E_\mathrm{S} \!\sum_{p=0}^P \!\hat{\sigma}^{2(i-1)}_{h:{n,p}} |\gamma_{p:n,m}|^2 \!\!,
\label{eq:variance_term_zeta_definition}
\end{eqnarray}
where the received signal estimate $\hat{y}^{(i)}_{x_m:n,p}$ after cancellation of the $m$-th soft symbol estimate is given by
\vspace{-1ex}
\begin{equation}
\label{eq:auxhvariancedatasoft}
\vspace{-1ex}
\hat{y}^{(i)}_{x_m:n,p} \triangleq \sum_{q \neq m}^{N-1} \gamma_{p:n,q} \hat{x}_{n,q}^{(i-1)}.
\end{equation}

In a similar fashion, the channel-centric replica can also be expressed as
\vspace{-2ex}
\begin{equation}
\label{eq:soft_IC_for_channel_coeff}
\vspace{-1ex}
\tilde{y}_{h:{p,n}}^{(i)} = y_n - \sum_{q \neq p}^P \hat{h}_{n,q}^{(i)} \cdot \hat{y}_{h:{n,q}}^{(i)},
\end{equation}
with the corresponding variance given by
\vspace{-1ex}
\begin{eqnarray}
\vspace{-1ex}
\tilde{\sigma}^{2(i)}_{h:{p,n}}\! \triangleq \!\sum_{q \neq p}^P\! \hat{\sigma}_{h:{n,q}}^{2(i-1)} |\hat{y}_{h:{n,q}}^{(i)}|^2 \! + \!\!\! \sum_{m=0}^{N-1}\!\! \hat{\sigma}^{2(i-1)}_{x:{n,m}} |\tilde{\gamma}_{h_p:{m,n}}^{(i)}|^2  \!+\! N_0 +\!\! \nonumber && \\
&&\hspace{-54ex} \sum_{q \neq p}^P\!\! \hat{\sigma}_{h:{n,q}}^{2(i)}\!\! \sum_{m=0}^{N-1}\!\! \hat{\sigma}^{2(i)}_{x:{n,m}} |\gamma_{q,nm}|^2 \!+ \!\sigma^2_h \sum_{m=0}^{N-1} \hat{\sigma}^{2(i)}_{x:{n,m}} |\gamma_{p:n,m}|^2 \!,
\label{eq:variance_term_zeta_definition_channel} 
\end{eqnarray}
where the channel-centric soft channel estimate $\hat{y}_{h:{n,p}}^{(i)}$ and the corresponding \ac{sIC} effective channel gain of the $p$-th path $\tilde{\gamma}_{h_p:{m,n}}^{(i)}$ are respectively defined as
\vspace{-1ex}
\begin{equation}
\label{eq:auxhvariancedata_and_channel}
\hat{y}_{h:{n,p}}^{(i)} \triangleq \sum_{m=0}^{N-1} \gamma_{p:n,m}\hat{x}_{n,m}^{(i-1)} \;\,\text{and}\;\,
\tilde{\gamma}_{h_p:{m,n}}^{(i)} \triangleq \sum_{q \neq p}^P \hat{h}_{n,q}^{(i-1)} \gamma_{q,nm}.
\end{equation}

\subsubsection{Belief Generation}

Beginning with the assumption that $G \cdot N \cdot P$ is a large enough scalar and that the individual estimation errors in $\hat{h}_{n,p}^{(i-1)}$ and $\hat{x}_{n,q}^{(i-1)}$ are uncorrelated, \ac{SGA} can be applied to generate the beliefs of all the symbols by substituting equation \eqref{eq:elementwise_final_input_output_relation} into \eqref{eq:soft_IC_process} to obtain the belief corresponding to the $m$-th symbol $x_m$ at the $n$-th factor node by combining the contributions of all signals in $\mathbf{y}$, excluding $y_n$, as
\vspace{-1ex}
\begin{equation}
\!\!\!\! p_{\text{x} | x_m} (\text{x} | x_m)\! =\! \prod_{q \neq n}^N\! p_{\text{y}_q | \text{x}_m}(\text{y}_q | x_m) \propto \text{exp} \bigg(\!\!\! - \!\frac{|\text{x} \!-\! \tilde{x}_{n,m}^{(i)}|^2}{\tilde{\sigma}_{\tilde{x}:{n,m}}^{2(i)}}\! \bigg),\!
\label{eq:PDF_extrinsic_belief}
\end{equation}
to extract the desired beliefs and their variances following
\begin{equation}
\label{eq:mean_and_var_of_extrinsic_belief}
\tilde{x}_{n,m}^{(i)}\!\! \triangleq \!\tilde{\sigma}_{\tilde{x}:{n,m}}^{2(i)}\!\! \sum_{q \neq n}^N \!\!\frac{\tilde{\gamma}_{x:{m,q}}^{*(i)} \tilde{y}_{x:{m,q}}^{(i)}}{\tilde{\sigma}_{x:{m,q}}^{2(i)}}\;\text{and}\;
\tilde{\sigma}_{\tilde{x}:{n,m}}^{2(i)}\!\! \triangleq \!\!\bigg(\! \sum_{q \neq n}^N \!\!\frac{|\tilde{\gamma}_{x:{m,q}}^{(i)}|^2}{\tilde{\sigma}_{x:{m,q}}^{2(i)}} \bigg)^{\!\!\!-1}\!\!\!\!.
\end{equation}

Similarly, the extrinsic beliefs of the channel gains can be modelled under \ac{SGA} by the approximate distribution
\vspace{-1ex}
\begin{equation}
p_{\text{h} | h_p} (\text{h} | h_p) \propto \text{exp} \bigg( - \frac{|\text{h} - \!\tilde{\;\mathrm{h}}_{n,p}^{(i)}|^2}{\tilde{\sigma}_{\!\tilde{\;\mathrm{h}}:{n,p}}^{2(i)}} \bigg)   
\label{eq:PDF_extrinsic_belief_for_channel},
\end{equation}
\vspace{-2ex}

\noindent to extract the desired beliefs and their variances following
\begin{equation}
\label{eq:mean_and_variance_of_extrinsic_belief_for_channel}
\!\tilde{\;\mathrm{h}}_{n,p}^{(i)}\! \triangleq \!\tilde{\sigma}_{\!\tilde{\;\mathrm{h}}:{n,p}}^{2(i)} \sum_{q \neq n}^N \frac{\hat{y}_{h:{qp}}^{*(i)} \tilde{y}_{h:{p,q}}^{(i)}}{\tilde{\sigma}_{h:{p,q}}^{2(i)}}\;\,\text{and}\;\,
\tilde{\sigma}_{\!\tilde{\;\mathrm{h}}:{n,p}}^{2(i)}\! \triangleq \!\!\bigg(\! \sum_{q \neq n}^N \frac{|\hat{y}_{h:{qp}}^{(i)}|^2}{\tilde{\sigma}_{h:{p,q}}^{2(i)}} \bigg)^{\!\!-1}\!\!\!\!\!.
\end{equation}

\floatstyle{spaceruled}% Select new float style
\restylefloat{algorithm}
\begin{algorithm}[t!]
    \caption{Proposed \ac{PBiGaBP}-based \ac{JCEDD} Technique for Oversampled \ac{AFDM} Systems in Doubly-Dispersive Channels}
    \label{alg:JCDE_PBiGaBP}
    \setlength{\baselineskip}{11pt}
    \textbf{Input:} demodulated receive signal $\mathbf{y}$, pilot symbols $\mathbf{x}_p$, \ac{DD} matrices $\mathbf{\Gamma}_p$, constellation power $E_\mathrm{S}$, noise variance $N_0$, channel variance per path $\sigma^2_h$ and damping factors $\beta_x$ and $\beta_h$. \\
    \textbf{Output:} decoded symbols $\mathbf{x}_d$ and channel estimates $\hat{h}_p, \forall p$.
    \vspace{-2ex} 
    \begin{algorithmic}[1]  
    \STATEx \hspace{-3.5ex}\hrulefill
    \STATEx \hspace{-3.5ex}\textbf{Initialization}
    \STATEx \hspace{-3.5ex} - Set iteration counter to $i=0$ and amplitudes $c_x = \sqrt{E_\mathrm{S}/2}$
    \STATEx \hspace{-3.5ex} - Fix pilots to $\hat{x}_{n,m}^{(i)} = [\mathbf{x}_p]_m$ and set corresponding variances
    \STATEx \hspace{-2ex} to $\hat{\sigma}^{2(i)}_{x:{n,m}} = 0, \forall n,m \in \mathcal{M}_p$
    \STATEx \hspace{-3.5ex} - Set initial data and channel estimates to $\hat{x}_{n,m}^{(0)} \!\!=\!\! 0$ and $\hat{h}_{n,p}^{(0)} \!\!=\!\! 0$
    \STATEx \hspace{-2.1ex} and corresponding variances to $\hat{\sigma}^{2(0)}_{x:{n,m}} = E_\mathrm{S}, \forall n,m \in \mathcal{M}_d$
    \STATEx \hspace{-2.1ex} and $\hat{\sigma}^{2(0)}_{h:{n,p}} = \sigma_h^2, \forall n, p$, respectively.
    \STATEx \hspace{-3.5ex}\hrulefill
    \STATEx \hspace{-3.5ex}\textbf{for} $i=1$ to $i_\text{max}$ \textbf{do}
    \STATEx \textbf{Channel Estimation}: $\forall n, p$
    \STATE Compute the variables $\hat{y}_{h:{n,p}}^{(i)}$ and $\tilde{\gamma}_{h_p:{m,n}}^{(i)}$ from eq.  \eqref{eq:auxhvariancedata_and_channel}.
    \vspace{-0.25ex}
    \STATE Compute soft signal $\tilde{y}_{h:{p,n}}^{(i)}$ and its corresponding variance $\tilde{\sigma}^{2(i)}_{h:{p,n}}$ from equation \eqref{eq:soft_IC_for_channel_coeff} and equation \eqref{eq:variance_term_zeta_definition_channel}, respectively.
    \vspace{-1.8ex}
    \STATE Compute extrinsic channel belief $\tilde{h}_{n,p}^{(i)}$ and its corresponding variance $\tilde{\sigma}_{\tilde{h}:{n,p}}^{2(i)}$ from equation \eqref{eq:mean_and_variance_of_extrinsic_belief_for_channel}.
    \vspace{-0.25ex}
    \STATE Compute denoised and damped channel estimate $\hat{h}_{n,p}^{(i)}$ and its corresponding variance $\hat{\sigma}_{h:{n,p}}^{2(i)}$ from equation \eqref{eq:Gaussian_denoiser_channel_damped} and equation \eqref{eq:Gaussian_denoiser_MSE_damped}, respectively.
    \STATEx \textbf{Data Detection}: $\forall n, m$
    \STATE Compute auxiliary variables $\tilde{\gamma}_{x:{m,n}}^{(i)}$ and $\hat{y}^{(i)}_{x_m:n,p}$ from equations \eqref{eq:effective_channel_gain_a_in_PDF} and \eqref{eq:auxhvariancedatasoft}, respectively.
    \STATE Compute soft signal $\tilde{y}_{x:{m,n}}^{(i)}$ and its corresponding variance $\tilde{\sigma}^{2(i)}_{x:{m,n}}$ from equation \eqref{eq:soft_IC_process} and equation \eqref{eq:variance_term_zeta_definition}, respectively.
    \STATE Compute extrinsic data belief $\tilde{x}_{n,m}^{(i)}$ and its corresponding variance $\tilde{\sigma}_{\tilde{x}:{n,m}}^{2(i)}$ from equation \eqref{eq:mean_and_var_of_extrinsic_belief}.
    \STATE Compute denoised and damped data estimate $\hat{x}_{n,m}^{(i)}$ and its corresponding variance $\sigma_{x:{n,m}}^{2(i)}$ from equations \eqref{eq:QPSK_denoiser} and \eqref{eq:QPSK_denoiser_damped} and equations \eqref{eq:MSE_x_m} and \eqref{eq:MSE_x_m_damped}, respectively.
    \STATEx \hspace{-3.5ex}\textbf{end for}
    \end{algorithmic}
\end{algorithm}
% \vspace{-0.4ex}

\subsubsection{Soft Replica Generation}

Finally, once again exploiting \ac{SGA} and explicitly defining $p_{\text{x} | x} \big(\text{x}|x;\tilde{x}_{n,m}^{(i)},\tilde{\sigma}_{\tilde{x}:{n,m}}^{2(i)}\big)$ and $p_{\text{h} | h} \big(\text{h}|h;\tilde{h}_{n,p}^{(i)},\tilde{\sigma}_{\tilde{h}:{n,p}}^{2(i)}\big)$ to be the likelihood functions of the data symbols and channel beliefs given in equations \eqref{eq:PDF_extrinsic_belief} and \eqref{eq:PDF_extrinsic_belief_for_channel}, respectively, the soft replicas of $x_m$ and $h_p$ can be obtained from the conditional expectation given the extrinsic beliefs as
\vspace{-1ex}
\begin{equation}
\vspace{-1ex}
\hat{x}_{n,m}^{(i)} =  \frac{\sum\limits_{x \in \mathcal{C}} x\cdot p_{\text{x} | x} \big(\text{x}|x;\tilde{x}_{n,m}^{(i)},\tilde{\sigma}_{\tilde{x}:{n,m}}^{2(i)}\big)\cdot p_{x}(x)}{\sum\limits_{x' \in \mathcal{C}} p_{\text{x} | x'} \big(\text{x}|x';\tilde{x}_{n,m}^{(i)},\tilde{\sigma}_{\tilde{x}:{n,m}}^{2(i)}\big)\cdot p_{x'}(x')},
\label{eq:Bayes_rule_data}
\end{equation}
\vspace{-1ex}
\begin{equation}
\vspace{-1ex}
\hat{h}_{n,p}^{(i)} = \frac{\int h\cdot p_{\text{h} | h} \big(\text{h}|h;\tilde{h}_{n,p}^{(i)},\tilde{\sigma}_{\tilde{h}:{n,p}}^{2(i)}\big)\cdot p_{h}(h)}{\int p_{\text{h} | h'} \big(\text{h}|h';\tilde{h}_{n,p}^{(i)},\tilde{\sigma}_{\tilde{h}:{n,p}}^{2(i)}\big)\cdot p_{h'}(h')}.
\label{eq:Bayes_rule_channel}
\end{equation}

Next, under the assumption that each element $x$ of the data vector is independently drawn with equal probability out of $Q$-points contained in the constellation $\mathcal{C}$ and the prior $p_x(x)$ is a Multinomial distribution of the $Q$-th order, the beliefs $\hat{x}_{n,m}^{(i)}$ can be explicitly computed for \ac{QPSK} modulated symbols via
\vspace{-1ex}
\begin{equation}
\vspace{-1ex}
\hat{x}_{n,m}^{(i)}\! =\! c_x\! \cdot\! \bigg(\! \text{tanh}\!\bigg[ 2c_x \frac{\Real{\tilde{x}_{n,m}^{(i)}}}{\tilde{\sigma}_{\tilde{x}:{n,m}}^{2(i)}} \bigg]\!\! +\! j\text{tanh}\!\bigg[ 2c_x \frac{\Imag{\tilde{x}_{n,m}^{(i)}}}{\tilde{\sigma}_{\tilde{x}:{n,m}}^{2(i)}} \bigg]\!\bigg),\!\!
\label{eq:QPSK_denoiser}
\end{equation}
where $c_x \triangleq \sqrt{E_\mathrm{S}/2}$ denotes the magnitude of the real and imaginary parts of the explicitly chosen \ac{QPSK} symbols.

To further improve convergence \cite{Su_TSP_2015}, the final output is computed by damping the result $\hat{x}_{n,m}^{(i)}$ from equation \eqref{eq:QPSK_denoiser} with a damping factor $0 < \beta_x < 1$ to yield
\vspace{-1ex}
\begin{equation}
    \vspace{-1ex}
\label{eq:QPSK_denoiser_damped}
\hat{x}_{n,m}^{(i)} = \beta_x \hat{x}_{n,m}^{(i)} + (1 - \beta_x) \hat{x}_{n,m}^{(i-1)},
\end{equation}
with its corresponding variance $\hat{\sigma}^{2(i)}_{x:{n,m}}$ also updated and damped following equation \eqref{eq:MSE_x_m} to give
\vspace{-1ex}
\begin{equation}
    \vspace{-1ex}
\label{eq:MSE_x_m_damped}
\hat{\sigma}^{2(i)}_{x:{n,m}} = \beta_x (E_\mathrm{S} - |\hat{x}_{n,m}^{(i)}|^2) + (1-\beta_x) \hat{\sigma}_{x:{n,m}}^{2(i-1)}.
\end{equation}

Similarly, notice that the likelihood in equation \eqref{eq:Bayes_rule_channel} is a Gaussian-product distribution, with the mean of $p_{h}(h)$ equal to zero \cite{Parker_TSP_2014}.
Therefore, the channel estimate belief can be straightforwardly computed and damped with a damping factor $0 < \beta_h < 1$ via
\vspace{-1ex}
\begin{equation}
\vspace{-1ex}
\label{eq:Gaussian_denoiser_channel_damped}
\hat{h}_{n,p}^{(i)} = \beta_h \frac{\sigma^2_h \tilde{h}_{n,p}^{(i)}}{\tilde{\sigma}_{\tilde{h}:{n,p}}^{2(i)} + \sigma^2_h} + (1-\beta_h) \hat{h}_{n,p}^{(i-1)},
\end{equation}
and
\begin{equation}
\label{eq:Gaussian_denoiser_MSE_damped}
\hat{\sigma}_{h:{n,p}}^{2(i)} = \beta_h \frac{\sigma^2_h \tilde{\sigma}_{\tilde{h}:{n,p}}^{2(i)}}{\tilde{\sigma}_{\tilde{h}:{n,p}}^{2(i)} + \sigma^2_h} + (1-\beta_h) \hat{\sigma}_{h:{n,p}}^{2(i-1)}.
\end{equation}

The proposed \ac{PBiGaBP}-based \ac{JCEDD} algorithm is summarized as a pseudocode in Algorithm \ref{alg:JCDE_PBiGaBP}.

\section{Simulation Results}

\subsection{Numerical Simulations}

\begin{figure}[b!]
  \centering
  \captionsetup[subfloat]{labelfont=small,textfont=small}
  \subfloat[BER Performance.]{{\includegraphics[width=0.95\columnwidth]{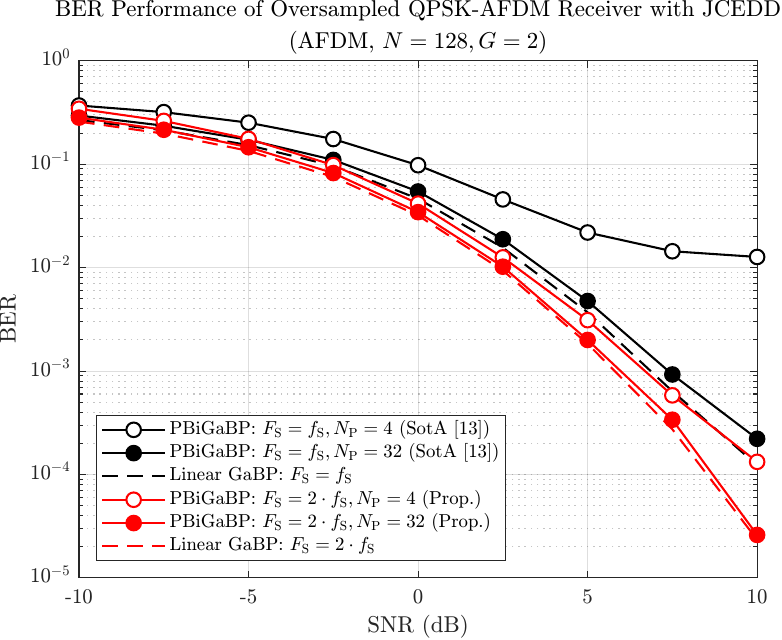}}}%
  \label{fig:BER_plot}
  \subfloat[NMSE Performance.]{{\includegraphics[width=0.95\columnwidth]{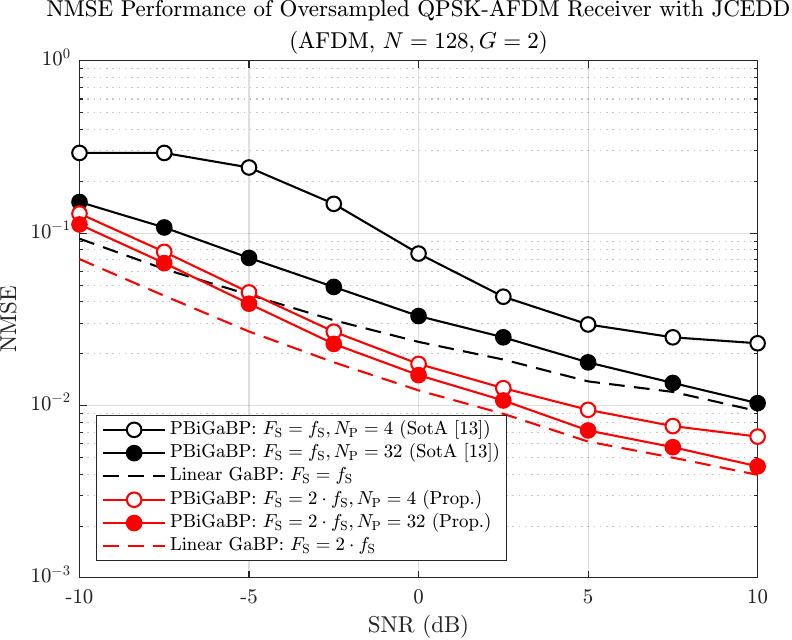}}}%
  \label{fig:NMSE_plot}
  %\vspace{-2ex}
  \caption{Performance of the proposed oversampled \ac{AFDM} receiver via the \ac{PBiGaBP} algorithm compared to the \ac{SotA} Nyquist receiver under a varying number of $N_\mathrm{P}$ pilots.}
  \label{fig:XXXM_Performance_plot}
  \vspace{-4ex}
\end{figure}

We consider an oversampled \ac{mmWave} \cite{WymeerschPIMRC2021,RanasingheARXIV2024} \ac{AFDM} system at $G = 2$ consisting of $N = 128$ subcarriers with $1$ \ac{LoS} path and $4$ \ac{NLoS} paths ($i.e., P = 4$) operating at $70 \, \text{GHz}$ with a bandwidth of $20 \, \text{MHz}$ using \ac{QPSK} symbols.
For the doubly-dispersive channel model, we consider a maximum unambiguous range of $75 \, \text{m}$ and a maximum unambiguous velocity of $602 \, \text{km/h}$, which gives us a maximum normalized delay index of $20$ and a maximum normalized digital Doppler shift index of $0.25$ (i.e., $\ell_\text{max} =  \frac{\tau_\text{max}}{T_\mathrm{S}} = 20$ and $f_\text{max} = N \nu_\text{max} T_\mathrm{S} = 0.25$). 
The path delays $\tau_p$ are then randomly generated using a uniform distribution across $[0,\tau_\text{max}]$ and Jakes' Doppler spectrum \cite{Bemani_TWC_2023} is used for the generation of the path Dopplers as $\nu_p = \nu_\text{max} \cos(\theta_p)$, where $\theta_p$ is uniformly distributed over $[-\pi,\pi]$.
\Ac{wlg}, $\text{P}_\mathrm{RC}$ in equation \eqref{eq:oversampled_TD_IO_relationship} is also simplified to a sinc filter, which corresponds to a \ac{RC} pulse with a rolloff factor 0.
Regarding the \ac{PBiGaBP} algorithm, we set $\beta_x, \beta_h = 0.3$, the maximum number of iterations $i_\text{max} = 40$, the constellation power $E_\mathrm{S}=1$ and the average channel power per path $\sigma_h^2 = 1$.

Proceeding by using the Linear \ac{GaBP}\footnote{An estimation method that uses \ac{GaBP} for both \ac{CE} (with full knowledge of symbols) and data decoding (with full knowledge of channel coefficients), via an \ac{LMMSE} initialization for \ac{CE}. This provides a lower bound on the performance for the proposed method in exchange for a large computational cost.} as a lower bound, we compare in Fig. \ref{fig:XXXM_Performance_plot} the communications and \ac{CE} performances, in terms of \ac{BER} and \ac{NMSE}, respectively, achieved by \ac{PBiGaBP} employed by the oversampled \ac{AFDM} receiver with $G = 2$.

The results show that the proposed oversampled \ac{PBiGaBP}-based \ac{JCEDD} with \ac{AFDM}  outperforms a system employing the same \ac{PBiGaBP} technique but using a \ac{SotA} Nyquist rate receiver, both in terms of \ac{BER} with a 2.5 dB gain and \ac{NMSE} with a 5 dB gain. 
In addition, we observe that the proposed receiver has an improved robustness to the change in the number of pilots with minimal performance degradation compared to the \ac{SotA} as seen from Fig. \ref{fig:XXXM_Performance_plot}, even when the pilot symbol percentage is changed from $25\%$ ($32$ pilots) to $3.125\%$ ($4$ pilots) due to the increase in factor nodes.

\subsection{Complexity Analysis}

Considering the typical floating point operations performed, the computational complexity of the proposed \ac{PBiGaBP} algorithm for oversampled systems is  $\mathcal{O}(N^2 G (P+1))$ which is linear on the number element-wise operations executed.
This is therefore of a much lower complexity than the compared Linear \ac{GaBP} method which incurs a total computational complexity of $\mathcal{O}(N^3)$ due to the costly matrix inversion carried out during the channel estimation procedure.

\section{Conclusion}

In this paper, we examined the \ac{JCEDD} performance of an oversampled \ac{AFDM} system in a doubly-dispersive surrounding.
By deriving a general \ac{TD} oversampling model for arbitrary waveforms, we present the oversampled \ac{AFDM} case as an example.
Then, we numerically verify that the proposed \ac{TD} oversampling mechanism yields significant performance gains, both in terms of \ac{BER} and \ac{NMSE} performance.
Finally, we reveal that oversampling significantly improves the robustness of the system against a varying number of pilots, given that the performance deviation is minimum even when the number of pilot symbols are reduced significantly.

\section*{Acknowledgement}

This work was supported in part by the RIE2020 Industry Alignment Fund-Industry Collaboration Projects (IAF-ICP) Funding Initiative, as well as cash and in-kind contribution from the industry partner(s).

% \newpage
\bibliographystyle{IEEEtran}

\end{document}